\documentstyle[aps,twocolumn,tighten,epsfig]{revtex}
\include{epsfig.sty}
\begin{document}

\draft
\title{ DIFFUSION OF CLASSICAL SOLITONS }

\author{Jacek Dziarmaga\thanks{E-mail: {\tt ufjacekd@thp1.if.uj.edu.pl},
present address: Institute of Physics, Jagiellonian University,
Reymonta 4, 30-059 Krak\'ow,Poland.}
    and Wojtek Zakrzewski\thanks{E-mail: {\tt W.J.Zakrzewski@durham.ac.uk}} }

\address{Department of Mathematical Sciences,
         University of Durham, South Road, Durham, DH1 3LE,
         United Kingdom}
\date{January 27, 1997}
\maketitle
\tighten

\begin{abstract}
{\bf We study the diffusion and deformation of classical solitons coupled
to
thermal noise. The diffusion coefficient for kinks in the $\phi^4$ theory
is predicted up to the second order in $kT$. The prediction is verified by 
numerical simulations.  Multiskyrmions in the vector O(3) sigma model are
studied within the same formalism. Thermal noise results in a diffusion on
the multisoliton collective coordinate space (moduli space).  There are
entropic forces which tend, for example, to bind pairs of solitons into
bi-solitonic  molecules. }
\end{abstract}

  Solitons or extended objects are an important ingredient in many
physical phenomena ranging from bioenergetics to superconductivity or
nonlinear optics, see e.g.\cite{davydov} for a review. Most of these
phenomena take place at a finite temperature and involve dissipation and
noise. Although it is sometimes possible to study transport properties of
solitons starting from first principles \cite{neto}, it seems that in most
cases one has to rely on some kind of an effective diffusive nonlinear
equation derived by expansion in derivatives of the order parameter, see
e.g. \cite{tsuneto} for an example.

  In the following we develop a perturbative expansion in powers of
temperature, which we use to predict the diffusion coefficient and
the deformation of kinks at finite temperatures. Our predictions are confirmed
by numerical simulations. Then we generalize the formalism to
multisolitons in the planar vector $O(3)$ sigma model. In the absence of
thermal noise solitons of the pure sigma model do not interact with each
other. Thus they provide a convenient setting for an ``in vitro" study of noise induced
entropic forces.

\section{Diffusion of kinks}

  Let us consider a dissipative version of the $\phi^{4}$ theory in one
spatial dimension, which is defined, in appropriate dimensionless units,
by the stochastic nonlinear field equation

\begin{equation}\label{model}
\Gamma\partial_{t}\phi=\partial_{x}^{2}\phi+2[1-\phi^{2}]\phi+\eta(t,x)
\;\;,
\end{equation}
where $\Gamma$ is a dissipation coefficient and $\eta(t,x)$ is a gaussian
white noise with correlations

\begin{eqnarray}\label{correlations}
&&<\eta(t,x)>=0 \;\;, \nonumber\\
&&<\eta(t_1,x_1)\eta(t_2,x_2)>=2kT\Gamma\delta(t_1-t_2)\delta(x_1-x_2)\;\;.
\end{eqnarray}
The system is coupled to an ideal heat bath at temperature $T$. At nonzero
temperature the field $\phi(t,x)$ performs a random walk in its
configuration space. In the absence of noise, at $T=0$, Eq.(\ref{model}) 
admits static kink solutions $\phi(t,x)=F(x)\equiv\tanh(x)$.  Antikinks
are given by $F(-x)$.

\paragraph*{\bf{Spectrum of kink excitations.}}

Small perturbations around the kink take the form $e^{-\gamma
t/\Gamma}u(x)$. Linearization of Eq.(\ref{model}) with respect to $u(x)$,
for $\eta(t,x)=0$, gives

\begin{equation}\label{geq}
\gamma\; u(x)=-\frac{d^{2}}{dx^{2}}\;u(x)+
            [4-\frac{6}{\cosh^{2}(x)}]\;u(x) \;\;.
\end{equation}
The eigenvalues and eigenstates can be tabulated as~\cite{jackiw}

\begin{eqnarray}\label{spectrum}
 \gamma,\;\;\;\;\;\;\;\;\;\;  && u(x)  \;\;,\nonumber\\
 0,\;\;\;\;\;\;\;\;\;\; && F'(x)=\frac{-1}{\cosh^{2}(x)}  \;\;,\nonumber\\
 3,\;\;\;\;\;\;\;\;\;\; &&
                B(x)\equiv\frac{\sinh(x)}{\cosh^{2}(x)}\;\;,\nonumber\\
 4+k^{2},\;\; && u_{k}(x)\equiv
 e^{ikx}[1+\frac{3ik\tanh(x)-3\tanh^{2}(x)}{1+k^2}]\;\;,
\end{eqnarray}
where $k$ is a real momentum. The zero mode ($\gamma=0$) is separated by a
gap from the first excited state (breather mode) of $\gamma=3$. The
continuum states are normalized so that

\begin{equation}\label{norm}
\int_{-\infty}^{+\infty} dx\; u_{k}^{*}(x) u_{k'}(x)=
2\pi\;\frac{4+k^2}{1+k^2}\;\delta(k-k')\equiv N(k)\;\delta(k-k') \;\;.
\end{equation}
In the following we will often denote the states (\ref{spectrum})
by $u_{a}(x)$ with the index $a=0$ reserved for the zero mode.

\paragraph*{\bf{Collective coordinates.}}

The field in the one kink sector of the theory (\ref{model}) can be
expanded in the complete set (\ref{spectrum})  as

\begin{equation}\label{expansion}
\phi(t,x)=F[x-\xi(t)]+\sum_{a\neq 0} A_{a}(t)\;u_{a}[x-\xi(t)] \;\;.
\end{equation}
Substitution of the above to Eq.(\ref{model}) and projection on the
orthogonal basis (\ref{spectrum}) gives a set of stochastic nonlinear
differential equations

\begin{eqnarray}\label{fullset}
&& \Gamma\;N_{0}\;\dot{\xi}(t)
   -\Gamma\;\dot{\xi}(t)\;\sum_{a\neq 0}A_{a}(t)\;M_{a0}
   +\sum_{b,c\neq 0} A_{b}(t)A_{c}(t)\;P_{bc0}+ \nonumber\\
&& \sum_{b,c,d\neq 0} A_{b}(t)A_{c}(t)A_{d}(t)\;R_{bcd0}
   =\eta_{0}(t) \;\;,        \nonumber \\
&& \Gamma\;N_{a}\;\dot{A}_{a}(t) 
   +\gamma_{a}\;N_{a}\;A_{a}(t) 
   -\Gamma\;\dot{\xi}(t)\;\sum_{b\neq 0}A_{b}(t)\;M_{ba}+ \nonumber \\
&& \sum_{b,c\neq 0} A_{b}(t)A_{c}(t)\;P_{bca}+
   \sum_{b,c,d\neq 0} A_{b}(t)A_{c}(t)A_{d}(t)\;R_{bcda}=\eta_{a}(t)
\;\;,
\end{eqnarray}
where $\dot{}\equiv d/dt$ and the coefficients are

\begin{eqnarray}\label{coefficients}
&& N_{a}=\int_{-\infty}^{+\infty}dx\; u_{a}(x)u_{a}^{*}(x)\;\;,\nonumber\\
&& M_{ab}=\int_{-\infty}^{+\infty}dx\;u'_{a}(x)u_{b}^{*}(x)\;\;,\nonumber\\
&& P_{abc}=6\;\int_{-\infty}^{+\infty}dx\; 
           F(x)\;u_{a}(x)u_{b}(x)u_{c}^{*}(x) \;\;,\nonumber\\
&& R_{abcd}=2\;\int_{-\infty}^{+\infty}dx\;
            u_{a}(x)u_{b}(x)u_{c}(x)u_{d}^{*}(x) \;\;.
\end{eqnarray}
The $\eta_{a}$'s result from the projections

\begin{equation}\label{projectednoise}
\eta_{a}(t)=\int_{-\infty}^{+\infty}dx\;\eta(t,x)u_{a}^{*}(x) \;\;.
\end{equation}
With the correlations (\ref{correlations}) and the orthogonality of the
basis (\ref{spectrum}) the correlations of the projected noises are

\begin{eqnarray}\label{projectedcorrelations}
&&<\eta_{a}(t)>=0 \;\;,\nonumber\\
&&<\eta_{a}^{*}(t_1)\eta_{b}(t_2)>=
  2kT\Gamma\;N_{a}\;\delta_{ab}\;\delta(t_1-t_2) \;\;.
\end{eqnarray}

\paragraph*{\bf{Diffusion coefficient to leading order in $\bf{kT}$.}}

  The natural parameter of expansion at low temperature is $\sqrt{kT}$. 
The projected noises, compare Eq.(\ref{projectedcorrelations}), are just
of this order. Let us make the customary rescaling
$\sqrt{kT}\rightarrow\varepsilon\sqrt{kT}$. Eqs.(\ref{fullset}) can be
expanded in powers of $\varepsilon$ and solved in power series of
$\varepsilon$ with $\varepsilon$ set to $1$ at the end of the calculation.
With the expansion of the collective coordinates, $\xi=\varepsilon
\xi^{(1)} +\varepsilon^2 \xi^{(2)}+\ldots$ and $A_{a}=\varepsilon
A_{a}^{(1)}+ \varepsilon^2 A_{a}^{(2)}+\ldots$, the equations
(\ref{fullset}) become, to the leading order in $\varepsilon$,

\begin{eqnarray}\label{leadingset}
&&\Gamma\;N_{0}\;\dot{\xi}^{(1)}(t)=\eta_{0}(t) \;\;,\nonumber \\
&&\Gamma\;N_{a}\;\dot{A}_{a}^{(1)}(t)
  +\gamma_{a}\;N_{a}\;A_{a}^{(1)}(t)=\eta_{a}(t) \;\;.
\end{eqnarray}
In this approximation the zero mode $\xi^{(1)}(t)$ and the excited modes
$A_{a}^{(1)}(t)$ are uncorrelated stochastic processes driven, respectively,
by their mutually uncorrelated projected noises.

  $\xi^{(1)}(t)$ is a Markovian Wiener process whose only nonvanishing
single connected correlation function is

\begin{equation}\label{xixi}
<\dot{\xi}^{(1)}(t)\dot{\xi}^{(1)}(t')>=
 \frac{2kT}{\Gamma N_{0}}\;\delta(t-t') \;\;,
\end{equation}
which is singular for $t\rightarrow t'$. According to the first of
Eqs.(\ref{leadingset}), the probability $P(t,\xi)$, that the kink random
walks to $\xi$ at the time $t$, satisfies the diffusion equation

\begin{equation}
\frac{\partial P}{\partial t}=D\;\frac{\partial^2 P}{\partial \xi^2} 
\end{equation}
with the diffusion coefficient $D=kT/\Gamma N_{0}=3kT/4\Gamma$.

 $A^{(1)}_a$'s are Ornstein-Uhlenbeck random noises with relaxation times
$\Gamma/\gamma_{a}$. They represent memory effects. If, for $t<0$, $T=0$ and
then the system is put in contact with a heat bath of temperature $T>0$,
for $t>0$ the correlations of $A^{(1)}$'s would grow according to

\begin{equation}\label{AA}
<[A^{(1)}_{a}(t)]^{*}\;A^{(1)}_{b}(t')>=
\delta_{ab}\;
\frac{kT\;e^{\frac{-\gamma_{a}|t-t'|}{\Gamma}}}{N_{a}\gamma_{a}}\;
[1-e^{\frac{-\gamma_{a}(t+t')}{\Gamma}}] \;\;,
\end{equation}
as can be deduced from a formal solution of the second of
Eqs.(\ref{leadingset}). All other single connected correlations of
$A^{(1)}$'s vanish. A given mode achieves the state of equilibrium with
the heat bath after its characteristic relaxation time. The longest
relaxation time is that of the breather mode.

\paragraph*{\bf{The next to leading order correction.}}

  Further terms in the expansion of collective coordinates can be
recursively worked out as

\begin{eqnarray}\label{terms}
\dot{\xi}^{(2)}(t)&=&
  \dot{\xi}^{(1)}(t) \sum_{a\neq 0} A_{a}^{(1)}(t) \frac{M_{a0}}{N_{0}}-
  \sum_{b,c\neq 0} A_{b}^{(1)}(t)A_{c}^{(1)}(t)
                   \frac{P_{bc0}}{\Gamma N_{0}} \;\;, \nonumber\\
A^{(2)}_{a}(t)&=&
  \sum_{b\neq 0}\frac{M_{ba}}{N_{a}}
  \int_{0}^{t} d\tau\; e^{-\frac{\gamma_{a}(t-\tau)}{\Gamma}}
                       \dot{\xi}^{(1)}(\tau)A_{b}^{(1)}(\tau)- \nonumber\\
&-&\sum_{b,c\neq 0}\frac{P_{bca}}{\Gamma N_{a}}
   \int_{0}^{t}d\tau\;e^{-\frac{\gamma_a(t-\tau)}{\Gamma}}
                      A_{b}^{(1)}(\tau)A_{c}^{(1)}(\tau) \;\;,\nonumber\\
\dot{\xi}^{(3)}(t)&=&
  \sum_{a\neq 0} [\dot{\xi}^{(1)}(t)A_{a}^{(2)}(t)+(1\leftrightarrow 2)]
                                        \frac{M_{a0}}{N_{0}}- \nonumber\\
&-&\sum_{b,c\neq 0} [A_{b}^{(1)}(t)A_{c}^{(2)}(t)+(1\leftrightarrow 2)]
                    \frac{P_{bc0}}{\Gamma N_{0}}-             \nonumber\\
&-&\sum_{b,c,d\neq 0} A_{b}^{(1)}(t)A_{c}^{(1)}(t)A_{d}^{(1)}(t)
                      \frac{R_{bcd0}}{\Gamma N_{0}} \;\;.
\end{eqnarray} 
In the thermodynamic equilibrium the excited modes develop nonzero
expectation values

\begin{equation}\label{A2}
<A_{a}^{(2)}(\infty)>=
-\frac{kT}{\gamma_{a}N_{a}}
[\frac{1}{2}P_{BBa}+
 \int_{-\infty}^{+\infty}dk\; \frac{P_{k,-k,a}}{\gamma(k)N(k)}] \;\;,
\end{equation}
where the index ``$B$" stands for the breather mode. In particular the
expectation value of the breather mode is negative,
$<A_{B}^{(2)}(\infty)>=-kT\frac{207\pi}{512}$. The kink gets thicker due
to the noise. At the same time the expectation value of $\phi(x,t)$, far from the kink, gets smaller,
$<\phi(\infty,\infty))>=1-\frac{3kT}{8}+O[(kT)^2]$. 

  The equilibrium correlations of $\dot{\xi}$, up to the next to the leading
order, are

\begin{eqnarray}
&&<\dot{\xi}(t)\dot{\xi}(t')>\approx
  \varepsilon^2 <\dot{\xi}^{(1)}(t)\dot{\xi}^{(1)}(t')>+ \nonumber\\
&&\varepsilon^4 <\dot{\xi}^{(2)}(t)\dot{\xi}^{(2)}(t')>+
  \varepsilon^4 [<\dot{\xi}^{(1)}(t)\dot{\xi}^{(3)}(t')>+
                 (1\leftrightarrow 3) ]=\nonumber\\
&&<\dot{\xi}^{(1)}(t)\dot{\xi}^{(1)}(t')>\times \nonumber\\
&&[ 1 + 
    \frac{3kT}{N_{0}^2}\sum_{a\neq 0} \frac{|M_{a0}|^2}{\gamma_a N_a} +
    \frac{2}{N_0}\sum_{a\neq 0}  M_{a0}<A_{a}^{(2)}(\infty)> ]+\ldots \;\;, 
\end{eqnarray}
where $\ldots$ denotes terms which are regular as $t\rightarrow t'$. In
the last equality we have already set $\varepsilon=1$. The integrals over
the continuous part of the spectrum and the summation over the breather mode
yielded $<\dot{\xi}(t)\dot{\xi}(t')>=\delta(t-t')\;
\frac{3kT}{2\Gamma}[1+kT\;1.8164]+O[(kT)^3]$.. The diffusion coefficient in
the equilibrium state, understood as an average over times much longer
than the relaxation time of the breather mode, has been found to be

\begin{equation}\label{D}
D(\infty)=\frac{3kT}{4\Gamma}[1+kT\;1.8164]+O[kT^3] \;\;.
\end{equation}

\paragraph*{\bf{Numerical experiment.}}

 The projection of the expansion (\ref{expansion}) onto $F'(x-y)$,

\begin{equation}
I(t,y)\equiv\int_{-\infty}^{+\infty}dx\; \phi(t,x)F'[x-y] \;\;,
\end{equation}
vanishes for $y=\xi(t)$. At moderate temperatures (small $A$'s) 
$I(t,y)>0$ for $y>\xi(t)$ and $I(t,y)<0$ for $y<\xi(t)$.Thus $\xi(t)$  can
be defined as the solution of $I[t,\xi(t)]=0$. This definition was
introduced in Ref.\cite{pla}; the coordinate defined in this way is a
position the kink would relax to if the noise were suddenly switched off.

  We performed numerical simulations of Eq.(\ref{model}) with $\Gamma=1$
for a range of temperatures. Fig.1 compares numerical results with the
perturbative prediction (\ref{D}). 

  In principle, for some periods of time, the solution of $I[t,\xi(t)]=0$
may happen to be not unique even for arbitrarily small $T>0$. There is
always small but finite density of thermally activated kink-antikink pairs.
Even if this happens for a given realization of the stochastic
process, one can still find two times $t_1<t_2$ such that the solution
is unique and choose at least one solution $\xi(t)$ valid for
any $t_1<t<t_2$; due to topology the equation has at least one solution for
any $t$. The ambiguity manifests
itself in the first of Eqs (\ref{fullset}) in the  coefficient of
$\dot{\xi}$, namely
$\Gamma\;[N_{0}-\sum_{a\neq 0}A_{a}(t)\;M_{a0}]$. The coefficient
vanishes and leads to ambiguity, iff $I'[t,\xi(t)]=0$ or a few zeros
meet at $\xi(t)$. At such a point, the trajectory splits into
several branches or a few branches merge into one. The perturbative
expansion overcomes this problem, it gives
an asymptotic approximation to the diffusion coefficient of the chosen
line $\xi(t)$.

\section{Diffusion of multisolitons.}

  A generalization of the ideas developed for a kink to a soliton in
higher dimensions is straightforward. Less trivial is the situation with
many solitons, which can overlap or pass through each other. Let us
consider a dissipative version of the $O(3)$ sigma model

\begin{equation}\label{sigmamodel}
\Gamma\;\partial_{t}\vec{M}=
\hat{\Pi}_{\vec{M}}[\nabla^2 \vec{M}+\vec{\eta}(t,\vec{x})] \;\;,
\end{equation}
where $\hat{\Pi}_{\vec{M}} \vec{A}=\vec{A}-\vec{M}(\vec{M}\vec{A})$ for
any vector $\vec{A}$ and the magnetization $\vec{M}$ is subject to the
constraint $\vec{M}\vec{M}=1$. The vector noise components are white
noises

\begin{eqnarray}\label{sigmacorrelations}
&&<\eta^{k}(t,\vec{x})>=0 \;\;, \nonumber\\
&&<\eta^{k}(t_1,\vec{x}_1)\eta^{l}(t_2,\vec{x}_2)>=
  2kT\;\Gamma\;\delta^{kl}\;\delta(t_1-t_2)\;
  \delta(\vec{x}_1-\vec{x}_2)
\end{eqnarray}
for any $k,l=1,2,3$.

 For a positive topological index $n$, in the absence of noise, the
solution of Eq.(\ref{sigmamodel}) relaxes to one out of the continuous
family of degenerate ground states~\cite{belavin}

\begin{eqnarray}\label{solution}
&&\vec{M}_{0}(\vec{x},c,a_i,b_i)=
  (\frac{W+W^*}{1+|W|^2},i\frac{W-W^*}{1+|W|^2},\frac{1-|W|^2}{1+|W|^2})
                                                        \;\;,\nonumber\\
&&W=c\frac{(z-a_1)\ldots (z-a_{n})}
          {(z-b_1)\ldots (z-b_{n})}  
\end{eqnarray}
parametrized by the complex parameters $a$'s, $b$'s and $c$;  $z=x_1-i
x_2$. There are $(4n+2)$ real parameters. $3$ of them can be removed by
fixing the boundary conditions at planar infinity and the remaining global
$U(1)$ symmetry~\cite{ward}.  After this reduction there are $(4n-1)$ real
parameters left, they parametrize a $(4n-1)$-dimensional real manifold
$M_{n}$, which is called the moduli space~\cite{manton}. We will denote the
real parameters by $\xi^{\alpha}$ with $\alpha=1,\ldots,(4n-1)$. The
static solution $\vec{M}_{0}(\vec{x},\xi)$ has $(4n-1)$ independent zero
modes $\partial \vec{M}_{0}(\vec{x},\xi)/\partial \xi^{\alpha}$. 

 As for kinks, compare Eq.(\ref{expansion}), the magnetization can be
expanded in excited states

\begin{equation}\label{sigmaexpansion}
\vec{M}(t,\vec{x})=\vec{M}_{0}[\vec{x},\xi(t)]\sqrt{1-K\sp2}\,+\,\vec K,
\end{equation}

where
\begin{equation}\label{sigmaexpansiona}
               \vec K= \sum_{a\neq 0}A_{a}(t)\vec{u}_{a}[\vec{x},\xi(t)] \;\;.
\end{equation}
Note that $\vec K\cdot M_0=0$.
The sigma model analogue of the first of Eqs. (\ref{leadingset}) is

\begin{equation}\label{sigmaset}
\Gamma\;g_{\alpha\beta}(\xi)\;\dot{\xi}^{\beta}=
\eta_{\alpha}(t,\xi) \;\;.
\end{equation}
where $g_{\alpha\beta}$ is a metric tensor on the moduli space

\begin{equation}
g_{\alpha\beta}(\xi)=\int d^2 x\;
\frac{\partial \vec{M}_{0}(\vec{x},\xi)}{\partial \xi^{\alpha}}
\frac{\partial \vec{M}_{0}(\vec{x},\xi)}{\partial \xi^{\beta}}
\;\;.
\end{equation}
Time correlations of the projected noises 

\begin{equation}\label{sigmanoises}
\eta_{\alpha}(t,\xi)=\int d^2 x\; 
\vec{\eta}(t,\vec{x})
\frac{\partial \vec{M}_{0}(\vec{x},\xi)}{\partial \xi^{\alpha}}
\end{equation}
can be obtained from the correlations (\ref{sigmacorrelations}) as

\begin{eqnarray}\label{sigmapc}
&&<\eta_{\alpha}(t,\xi)>=0 \;\;,\nonumber\\
&&<\eta_{\alpha}(t_1,\xi)\eta_{\beta}(t_2,\xi)>=
  2kT\;\Gamma\;g_{\alpha\beta}(\xi)\;\delta(t_1-t_2) \;\;.
\end{eqnarray}
The equation (\ref{sigmaset}) and the correlations (\ref{sigmapc}) lead to
the following diffusion equation on the moduli space

\begin{equation}\label{sigmadiffusion}
\frac{\partial P(t,\xi)}{\partial t}=
D_{\sigma}\; \frac{\partial}{\partial\xi^{\alpha}}\;
             \sqrt{g(\xi)}\;g^{\alpha\beta}(\xi)\;
             \frac{\partial}{\partial\xi^{\beta}}\;
             \frac{P(t,\xi)}{\sqrt{g(\xi)}} \;\;.
\end{equation}
$g(\xi)=det[g_{\alpha\beta}(\xi)]$,
$g^{\alpha\beta}=(g^{-1})_{\alpha\beta}$ and $D_{\sigma}=kT/\Gamma$ is a
diffusion coefficient. $P(t,\xi)$ is the probability of finding the
multisoliton at the point $\xi$ on the moduli space, such that the average
of any function $F(\xi)$ is given by $<F>=\int_{M_n} d\xi\; P(t,\xi)
F(\xi)$. If $\xi$ are restricted to a compact area of the moduli
space, the probability would finally relax to the equilibrium distribution
$P_{eq}(\xi)\sim\sqrt{g(\xi)}$. Interpreting this distribution as a
Boltzman distribution, we can identify the entropic potential on the
moduli space as

\begin{equation}
-TS(\xi)=-kT\;\ln\sqrt{g(\xi)} \;\;,
\end{equation}
which is not quite unexpected as $\sqrt{g(\xi)}$ is a surface element on
$M_n$.

\paragraph*{Example.}

 To give a clear but nontrivial example of entropic forces, let us
restrict our attention to the subspace of $M_2$ defined by $W=\frac{B^2}{z^2-A^2}$ with
real $A,B$. This space represents states of two solitons of the overall
size $B$. For $A>0$ ($A<0$) they are located on the real (imaginary) axis
at the points $\stackrel{+}{-}A$ ($\;\stackrel{+}{-}iA$). The sigma model
is scale invariant, so it is not surprising that $g(A,B)=g(A/B)$, compare
Eqs. (10-12) in Ref.\cite{ward},

\begin{equation}\label{Peq}
\sqrt{g(A/B)}=
  \sqrt{ 8  
         [\frac{E}{2}(K-\frac{E}{2})-(\frac{K-E}{2}\tan{\kappa})^2] 
         \sin{2\kappa} }\;\;,
\end{equation}
where $\kappa=\arctan(B/A)$ and $K=K[\cos(\kappa)]$, $E=E[\cos(\kappa)]$
are complete
elliptic integrals of the first and second kind respectively.  The
function $\sqrt{g}$ has a unique global maximum at $A/B\approx 1.26$,
compare Fig.2. For a definite scale $B$, which can be fixed by residual
interactions inducing an effective potential on the moduli space~\cite{jd}
such as $L-S$ coupling, Zeeman energy and a higher derivative (Skyrme)
term, the entropic forces tend to bind the two solitons $Sk$ into a
molecule $Sk_2$. At short distance there is a logarithmic repulsion.

\section*{Conclusions.}

  The numerical simulations confirm our method of expansion in powers of
$\sqrt{kT}$, when applied to kinks. We believe the same expansion is valid
for multisolitons in the sigma model. Even in the pure sigma model, in
which the solitons do not interact by potential forces, the thermal noise
induces entropic forces. In particular, two solitons attract each other
at large distances and repel at short distances, and so
 the entropic forces tend
to bind them into bi-solitonic  molecules.

\acknowledgements

J.D. would like to thank Nick Manton for an inspiring comment.  J.D. was
supported by a UK PPARC grant.

\centerline{\epsfbox{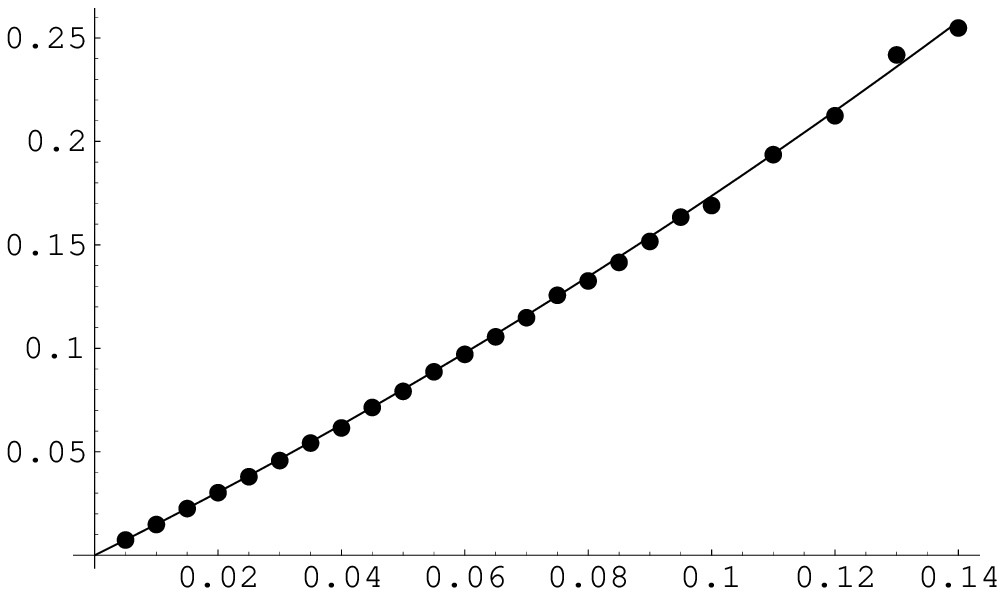}}

FIGURE 1. The equilibrium diffusion coefficient $D(\infty)$ for a kink as
a function of temperature $kT$. Solid line - perturbative prediction, see
Eq.(\ref{D}); dots - results from numerical simulation.
 
\centerline{\epsfbox{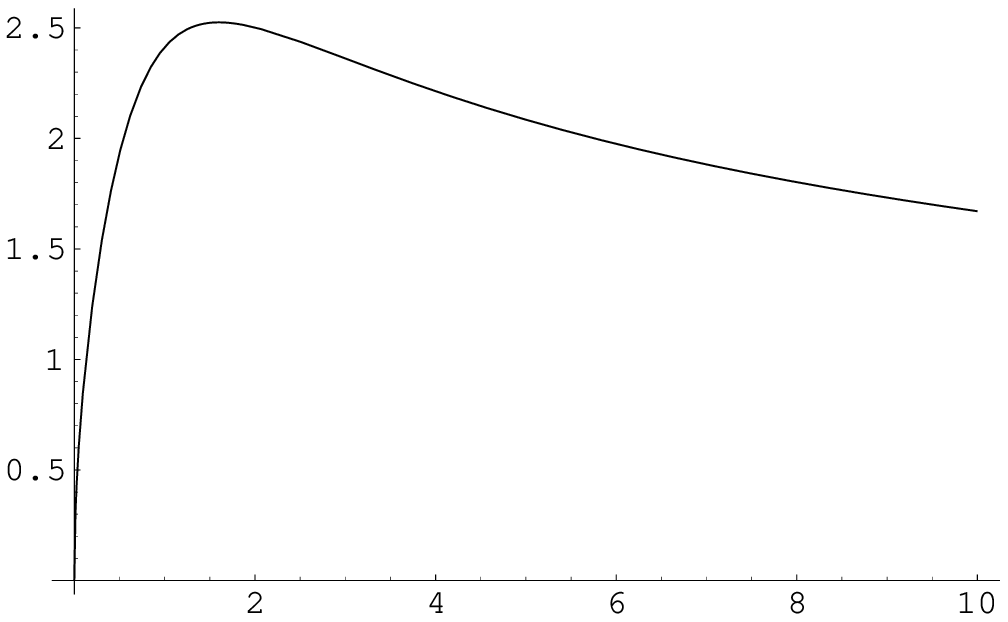}}
 
FIGURE 2. The probability distribution (\ref{Peq}) to find two skyrmions
separated by the distance A as a function of $A/B$. 

\end{document}